\documentclass[12pt]{article}
\usepackage{epsfig}
\begin{document}
\baselineskip=15pt
\begin{titlepage}
\setcounter{page}{0}
\begin{center}
\vspace*{5mm}
{\Large \bf Cosmology with a Variable Chaplygin Gas}\\
\vspace{15mm}
{\large Zong-Kuan Guo$^{1}$
 \footnote{e-mail address: guozk@phys.kindai.ac.jp}
 and Yuan-Zhong Zhang$^{2,3}$}\\
\vspace{10mm}
 {\it
 $^1$Department of Physics, Kinki University, Higashi-Osaka,
     Osaka 577-8502, Japan \\
 $^2$CCAST (World Lab.), P.O. Box 8730, Beijing 100080, China \\
 $^3$Institute of Theoretical Physics, Chinese Academy of
     Sciences, P.O. Box 2735, Beijing 100080, China}
\end{center}
\vspace{20mm}
\centerline{\large \bf Abstract}
 {We consider a new generalized Chaplygin gas model that
includes the original Chaplygin gas model as a special case. In
such a model the generalized Chaplygin gas evolves as from dust to
quiessence or phantom. We show that the background evolution for
the model is equivalent to that for a coupled dark energy model
with dark matter. The constraints from the current type Ia
supernova data favour a phantom-like Chaplygin gas model.} 
\vspace{2mm}
\begin{flushleft}
PACS number(s): 98.80.Es, 98.80.Cq
\end{flushleft}
\end{titlepage}

\section{Introduction}

Recent observations of type Ia supernovae (SNe Ia) suggest
that the expansion of the universe is accelerating and that
two-thirds of the total energy density exists in a dark
energy component with negative pressure~\cite{AGR}.
In addition, measurements of the cosmic microwave background
(CMB)~\cite{DNS} and the galaxy power spectrum~\cite{MT}
also indicate the existence of the dark energy.
The simplest candidate for the dark energy is a cosmological
constant $\Lambda$, which has pressure
$P_\Lambda=-\rho_\Lambda$. Specifically, a reliable model
should explain why the present amount of the dark
energy is so small compared with the fundamental scale
(fine-tuning problem) and why it is comparable with the
critical density today (coincidence problem).
The cosmological constant suffers from both these problems.
One possible approach to constructing a viable model for
dark energy is to associate it with a slowly evolving and
spatially homogeneous scalar field $\phi$, called
``quintessence''~\cite{RP,ZWS}. Such a model
for a broad class of potentials can give the energy density
converging to its present value for a wide set of initial
conditions in the past and possess tracker behavior
(see, e.g.,~\cite{VS00} for reviews with more complete
lists of references).

However, neither dark matter nor dark energy has laboratory
evidence for its existence directly. In this sense, our
cosmology depends on two untested entities. It would be nice
if a unified dark matter/energy (UDME) scenario can be found
in which these two dark components are different
manifestations of a single cosmic fluid~\cite{pad,CTC}.
An attractive feature of these models is that such an
approach naturally solves, at least phenomenologically, the
coincidence problem. As a candidate of the UDME scenarios,
the Chaplygin gas model was recently was proposed~\cite{KMP}.
The Chaplygin gas is characterized by an exotic equation
of state $P=-A/\rho$, where $A$ is a positive constant.
Such equation of state leads to a component which behaves
as dust at early stage and as cosmological constant at
later stage. The Chaplygin gas
emerges from the dynamics of a generalized $d$-brane in a
$(d+1,1)$ spacetime and can be described by a
complex scalar field whose action can be written as a generalized
Born-Infeld action~\cite{BBS1}.
The model parameters were constrained using
various cosmological observations, such as SN Ia data~\cite{MOW},
CMB experiments~\cite{BBS2} and other observational
data~\cite{DAJ}. The Chaplygin gas model has been extensively
studied in the literature~\cite{PFG}.

Recently, there are some indications that a strongly negative
equation of state, $w \le -1$, may give a good fit~\cite{MMOT}.
Here we propose a new generalized Chaplygin gas model in which the
Chaplygin gas can act like either quiessence with $w > -1$ or
phantom with $w < -1$ at low densities. Such a generalized model
may formally be derived from a Born-Infled Lagrangian density for
a scalar field. Alternatively, the same background evolution may
arise if there is an interaction between dark energy and dark
matter. We analyze its cosmological consequences and then
constrain the parameters associated with the generalized Chaplygin
gas using the recent SN Ia data.

\section{Model}

Let us now consider a Born-Infeld Lagrangian~\cite{SEN}
\begin{equation}
{\cal L}_{\mathrm{BI}}=V(\phi)
\sqrt{1+g^{\mu\nu}\partial_\mu\phi\partial_\nu\phi}\,,
\end{equation}
Where $V(\phi)$ is the scalar potential. In a spatially flat
Friedmann-Robertson-Walker (FRW) universe, the energy density and
the pressure are given by
$\rho = V(\phi)(1-\dot{\phi}^2)^{-1/2}$ and
$P = -V(\phi)(1-\dot{\phi}^2)^{1/2}$, respectively.
The corresponding equation of state is formally given by
\begin{equation}
\label{es}
P = -\frac{V^2(\phi)}{\rho}\,.
\end{equation}
If one rewrites the self-interaction potential as a function
of the cosmic scale factor: $V^2(\phi)=A(a)$~\cite{CJ}, then,
by inserting Eq.~(\ref{es}) into the the energy conservation
equation, $d\rho/d\ln a = -3(\rho+P)$, one finds
that the energy density evolves as
\begin{equation}
\label{ide}
\rho = a^{-3}\left[6\int A(a)a^5da+B\right]^{1/2},
\end{equation}
where $B$ is an integration constant. Given a function $A(a)$,
Eq.~(\ref{ide}) allows us to obtain a solution $\rho(a)$ in
principle. Following Refs.~\cite{CJ} we set $A(a)=A_0 a^{-n}$
where $A_0$ and $n$ are constants, and where we take $A_0>0$
and $n<4$. By taking explicitly the
integral (\ref{ide}) it follows that
\begin{equation}
\label{de}
\rho = \left[\frac{6}{6-n}\frac{A_0}{a^n}
 +\frac{B}{a^6}\right]^{1/2}\,.
\end{equation}
We find that $n=0$ corresponds the original Chaplygin gas model
which interpolates between a universe dominated by dust and a
De Sitter one.
Compared to the original Chaplygin gas,
in this generalized model (called as variable Chaplygin gas)
the universe tends to be a quiessence-dominated ($n>0$)~\cite{HM}
or phantom-dominated one ($n<0$)~\cite{RRC} with
constant equation of state parameter $w=-1+n/6$.
The first term on the right hand side of Eq.~(\ref{de}) is
initially negligible so that the expression (\ref{de})
can approximately be written as $\rho \sim a^{-3}$,
which corresponds to a universe dominated by dust-like
matter. Once the first term dominates, it causes the universe
to accelerate. In this case we find $a \sim t^{4/n}$ so that
the expansion is accelerated for $n<4$.
Defining
\begin{equation}
B_s \equiv \frac{B}{6A_0/(6-n)+B},
\end{equation}
In a flat FRW universe the Hubble parameter is now given by
\begin{equation}
\label{bi}
H(z)=H_0\left[B_s(1+z)^6
 +(1-B_s)(1+z)^n\right]^{1/4},
\end{equation}
where $z=1/a-1$ is redshift and $H_0$ is the present value
of the Hubble parameter. There are two free parameters in
this model, $B_s$ and $n$.

It is easy shown that the background evolution for the variable
Chaplygin gas model is equivalent to that for an interaction model
between the dark matter and the dark energy with
$w=-1+n/6$~\cite{LA}. Assuming the scaling behaviour for the dark
energy density and the dark matter density, $\rho_x \propto \rho_m
a^{6-n}$, in a flat FRW universe it is straightforward to get
\begin{equation}
\label{ia}
\rho_m+\rho_x=\rho_0\left[(1-\Omega_{m0})a^{-n}
 + \Omega_{m0}a^{-6}\right]^{1/2},
\end{equation}
where $\rho_0$ is the critical density and $\Omega_{m0}$ is the
matter density parameter. If the coupled system can be written
as $\dot{\rho}_m + 3H\rho_m =Q$ and
$\dot{\rho}_x + nH\rho_x/2 =-Q$,
such scaling solutions follow from an
interaction characterized by~\cite{ZP}
\begin{equation}
Q = -3 H \frac{1-n/6}{1+\rho_m/\rho_x} \rho_m,
\end{equation}
which indicates that there is a continuous transfer of energy
from the dark matter component to the dark energy for $n<4$.
Comparing Eq.~(\ref{ia}) with Eq.~(\ref{bi}) we see that $B_s$
can be interpreted as an effective matter density.
In the interaction scenario, constraints on
the model parameters ($\delta_0$, $w_{\phi}$, $\Omega_{m0}$)
from the SN data have been derived in Ref.~\cite{maj04}.
The variable Chaplygin gas model corresponds to an interaction
case with $\delta_0 = 3 (1-\Omega_{m0}) w_{\phi}$, which
corresponds to a line on the $(\delta_0,w_{\phi})$ plane
given a value of $\Omega_{m0}$. Form Fig.~6
in Ref.~\cite{maj04}, we see that our model is
consistent with their constrains on the interaction model.

\section{SN Ia Constraints}

We now consider constraints on the model through a statistical
analysis involving the most recent SN Ia data, as provided
recently by Riess {\it et al.}~\cite{AR}.
The total sample presented in Ref.~\cite{AR} consists of
186 events distributed over the redshift interval
$0.01 \le z \le 1.7$ and constitutes the compilation of the best
observations made so far by the two supernova search teams
plus 16 new events observed by the Hubble Space Telescope.
This total data set was divided into gold and silver
subsets. Here we will consider only the 157 events that
constitute the so-called gold sample.

\begin{figure}
\begin{center}
\includegraphics[width=8cm]{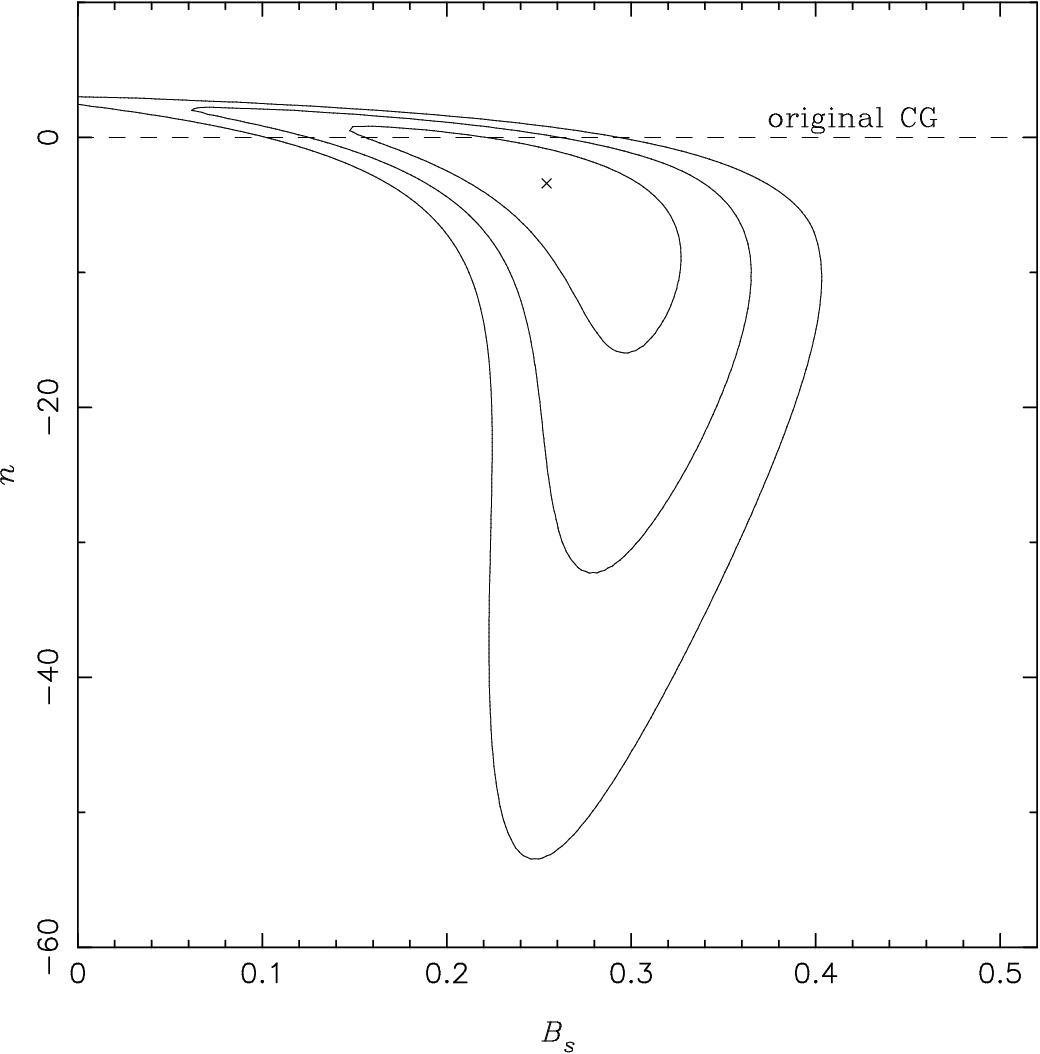}
\caption{Probability contours at 68.3\%, 95.4\% and 99.7\%
confidence levels for $B_s$ versus $n$ in the variable Chaplygin
gas model from the gold sample of 157 SN Ia data.
The dashed line represents the original Chaplygin gas model
with $n=0$. The best fit happens at $Bs=0.25$ and $n=-3.4$.}
\label{fig:chaplygin}
\end{center}
\end{figure}

The parameters in the model are determined by minimizing
\begin{equation}
\chi^2(H_0,B_s,n)=\sum_i\frac{[\mu_\mathrm{obs}
 (z_i)-\mu_\mathrm{mod}(z_i;H_0,B_s,n)]^2}{\sigma_i^2},
\end{equation}
where $\sigma_i$ is the total uncertainty in the observation,
the distance modulus $\mu(z_i)$ is
\begin{equation}
\mu(z_i)=5\log_{10}\frac{d_L(z_i)}{\mathrm{Mpc}}+25,
\end{equation}
and the luminosity distance in the spatially flat FRW model
with variable Chaplygin gas is given by
\begin{eqnarray}
d_L = cH_0^{-1}(1+z)\int_0^zdz \left[B_s(1+z)^6 +
(1-B_s)(1+z)^n\right]^{-1/4}.
\end{eqnarray}
To determine the likelihood of the parameters $B_s$ and
$n$, we marginalize the likelihood function
$L=\exp (-\chi^2/2)$ over $H_0$. We adopt a Gaussian prior
$H_0=72\pm 8$km s$^{-1}$ Mpc$^{-1}$ from the Hubble
Space Telescope Key Project~\cite{freedman}.
The results of our analysis are displayed in
Fig.~\ref{fig:chaplygin}. The best fit of the model gives
that $B_s = 0.25$ and $n = -3.4$ with $\chi^2 = 174.54$.
The three contours correspond to 68.3\%, 95.4\% and 99.73\%
confidence levels, respectively.
We can see that current SN Ia constraints fovour
a phantom-like Chaplygin gas model.

The age of the universe, $t(z)$, and the deceleration
parameter, $q(z)$, are given by
\begin{eqnarray}
t(z) &=& \int_{z}^{\infty}\frac{dx}{(1+x)H(x)}\,, \\
q(z) &=& \frac{d\ln H(z)}{d\ln (1+z)}-1.
\end{eqnarray}

\begin{figure}
\begin{center}
\includegraphics[width=8cm]{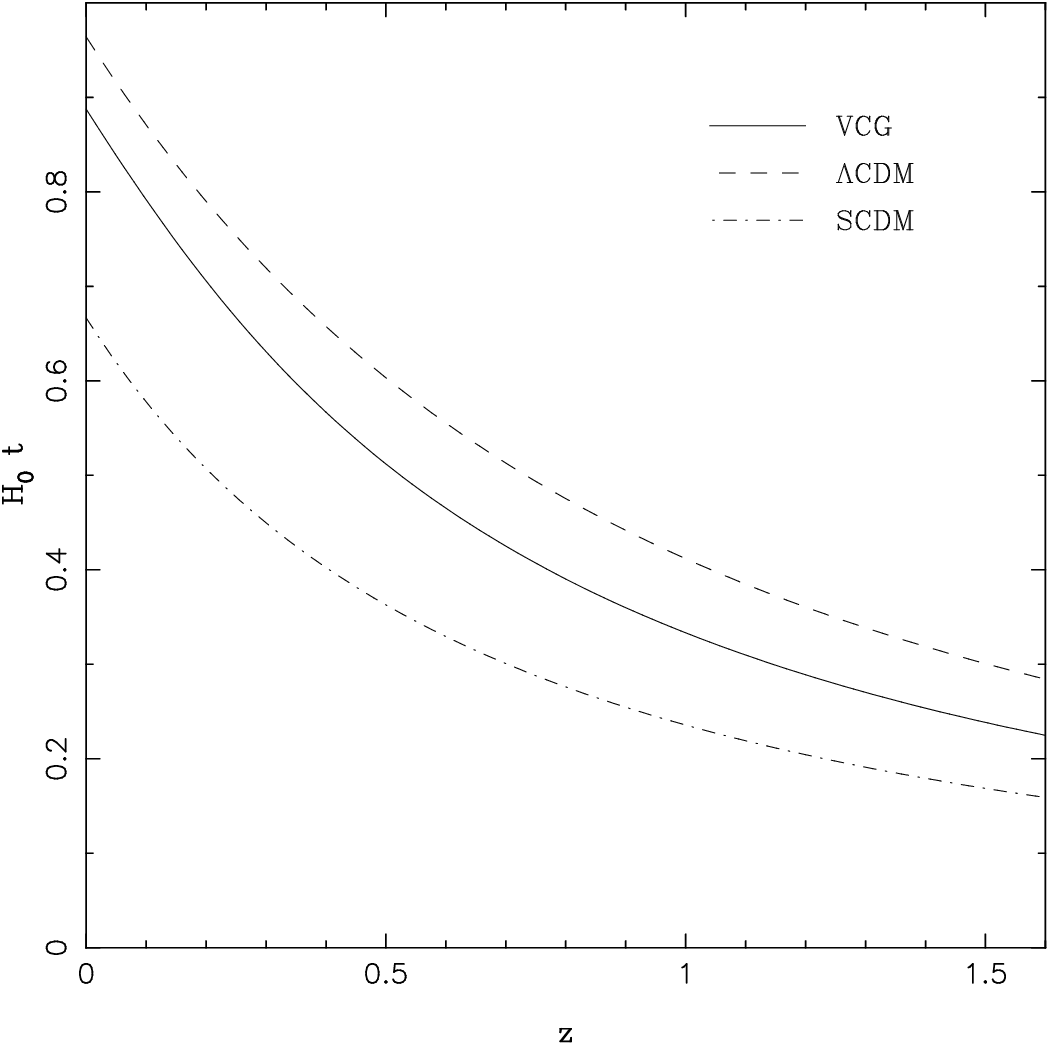}
\caption{The age of the universe, $H_0t(z)$, is shown as a
function of the redshift for the variable Chaplygin gas
model with $Bs=0.25$ and $n=-3.4$ (solid line), $\Lambda$CDM
with $\Omega_{m0}=0.3$ and $\Omega_{\Lambda}=0.7$
(dashed line) and SCDM (dot-dashed line).}
\label{fig:age}
\end{center}
\end{figure}

Fig.~\ref{fig:age} shows the evolution of the age of the universe
with redshift. We find that the best-fit age of the universe today
is $t_0=12.3$ Gyrs if the Hubble parameter is taken to be $H_0=72$
km s$^{-1}$ Mpc$^{-1}$~\cite{freedman}, which is slightly lower
than the age of a $\Lambda$CDM universe, $t_0=13.4$ Gyrs.
This age estimate is consistent with the results,
$t_0=12.5\pm2.5$ Gyrs at 95\% confidence level, from the oldest
globular clusters~\cite{marchi}.

\begin{figure}
\begin{center}
\includegraphics[width=8cm]{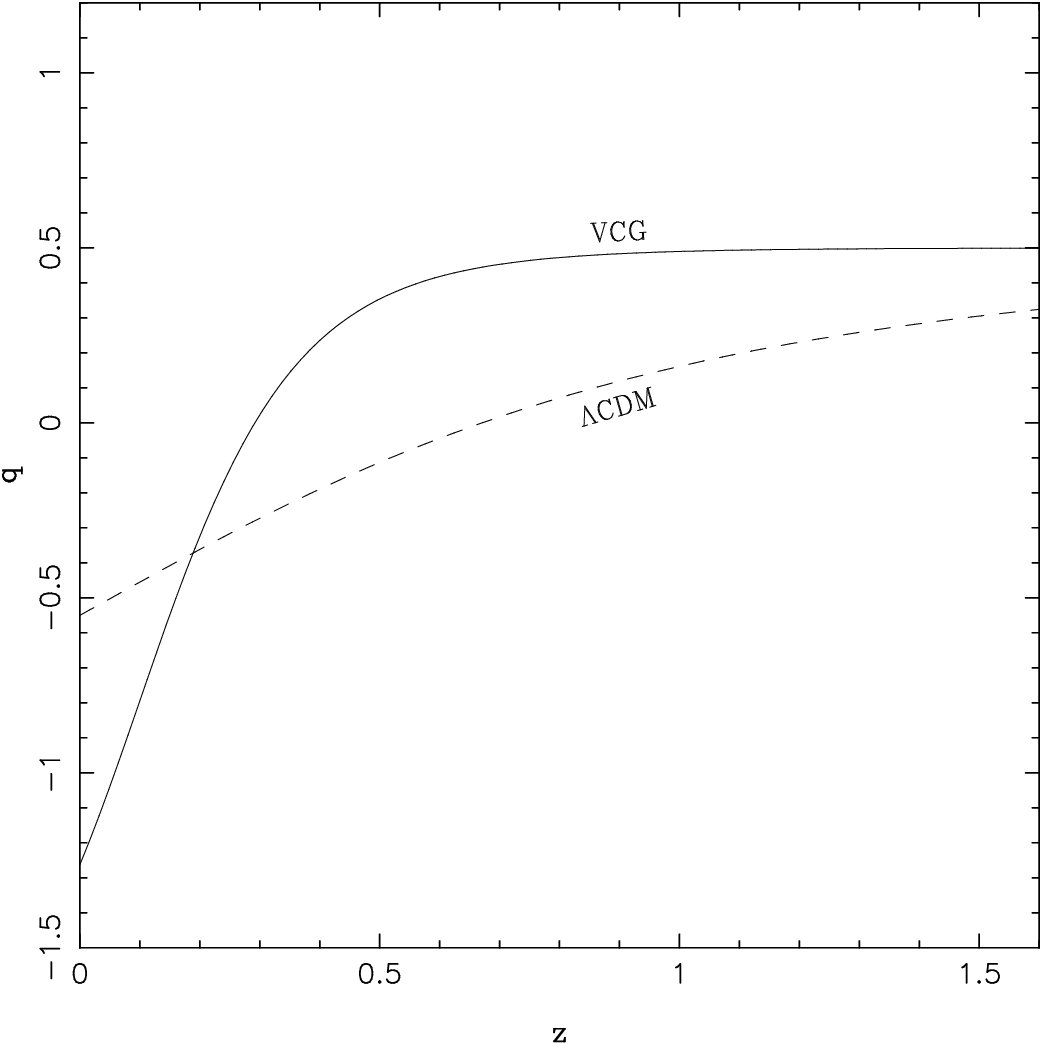}
\caption{Evolution of the deceleration of the universe,
$q(z)$ with redshift for the variable Chaplygin gas
model with $Bs=0.25$ and $n=-3.4$ (solid line) and
$\Lambda$CDM model with $\Omega_{m0}=0.3$ and
$\Omega_{\Lambda}=0.7$ (dashed line).}
\label{fig:qz}
\end{center}
\end{figure}

Fig.~\ref{fig:qz} shows the evolution of the deceleration
parameter with redshift. We find that the behaviour of the
deceleration parameter for the best-fit universe is quite
different from that in the $\Lambda$CDM cosmology.
The present value of the best-fit deceleration parameter,
$q_0 = -1.26$, is significantly lower than $q_0 = -0.55$ for
the $\Lambda$CDM model with $\Omega_{m0}=0.3$.
Furthermore, the rise of $q(z)$ with redshift is much steeper
in the case of the best-fit model, with the result that the
universe begins to accelerate at a comparatively lower
redshift $z=0.3$ (compared with $z=0.7$ for $\Lambda$CDM).

\section{Conclusions}

In this paper, we have considered a new generalized Chaplygin gas
model. In this scenario, the variable Chaplygin gas can drive the
universe from a non-relativistic matter dominated phase to an
accelerated expansion phase, behaving like dust-like matter in
early times and as quiessence/phantom in a recent epoch. We have
shown that the variable Chaplygin gas model is equivalent to an
interaction model between dark energy and dark matter in the sense
of the background evolution. Cosmic late-time acceleration implies
that there exists a continuous transfer of energy from the dark
matter component to the dark energy one. We constrained the
parameters associated with the variable Chaplygin gas using the
Gold SN sample. We find that the constrains from the SN data
favour a phantom-like Chaplygin gas model. This model deserves
further investigation as a viable cosmological model.

\section*{Acknowledgements}
The work of Z.K.G was supported in part by the Grant-in-Aid for
Scientific Research Fund of the JSPS Nos. 16540250 and 06042.
Y.Z.Z was partially supported by National Basic Research Program
of China under Grant No. 2003CB716300 and by NNSFC under Grant No.
90403032.

\end{document}